\documentclass[aps,pra,notitlepage,reprint,superscriptaddress,nofootinbib,longbibliography]{revtex4-1}

\usepackage{amssymb}
\usepackage{amsfonts}
\usepackage{comment}
\usepackage{graphicx}
\usepackage{dcolumn}
\usepackage{bm}
\usepackage{amsmath}
\usepackage{upgreek}
\usepackage[colorlinks,linkcolor=red,citecolor=blue,urlcolor=red]{hyperref}
\usepackage[utf8]{inputenc}
\usepackage[T1]{fontenc}
\usepackage{mathptmx}
\usepackage{ulem}
\usepackage[colorinlistoftodos]{todonotes}
\usepackage[mathlines]{lineno}
\renewcommand{\t}[1]{\mathrm{#1}}

\newcommand{\tunderbrace}[1]{\underbrace{\textstyle#1}}

\newcommand\numberthis{\addtocounter{equation}{1}\tag{\theequation}}

\newcommand{\nocontentsline}[3]{}
\let\origcontentsline\addcontentsline
\newcommand\stoptoc{\let\addcontentsline\nocontentsline}
\newcommand\resumetoc{\let\addcontentsline\origcontentsline}

\AtBeginDocument{%
	\newwrite\bibnotes
	\def\bibnotesext{Notes.bib}
	\immediate\openout\bibnotes=\jobname\bibnotesext
	\immediate\write\bibnotes{@CONTROL{REVTEX41Control}}
	\immediate\write\bibnotes{@CONTROL{%
			apsrev41Control,author="08",editor="1",pages="1",title="0",year="1"}}
	\if@filesw
	\immediate\write\@auxout{\string\citation{apsrev41Control}}%
	\fi
}

\usepackage{longtable}
\usepackage{ragged2e}

\begin{document}
	
	\title{Imaging-based Quantum Optomechanics}

\author{C. M. Pluchar}
\affiliation{Wyant College of Optical Sciences, University of Arizona, Tucson, AZ 85721, USA}
\author{W. He}
\affiliation{Wyant College of Optical Sciences, University of Arizona, Tucson, AZ 85721, USA}
\author{J. Manley}
\affiliation{Wyant College of Optical Sciences, University of Arizona, Tucson, AZ 85721, USA}
\author{N. Deshler}
\affiliation{Wyant College of Optical Sciences, University of Arizona, Tucson, AZ 85721, USA}
\author{S. Guha}
\affiliation{Wyant College of Optical Sciences, University of Arizona, Tucson, AZ 85721, USA}
\author{D. J. Wilson}
\affiliation{Wyant College of Optical Sciences, University of Arizona, Tucson, AZ 85721, USA}

\date{\today}
\begin{abstract}
	In active imaging protocols, information about an object is encoded into the spatial mode of a scattered photon.  \textcolor{black}{Recently the quantum limits of active imaging have been explored with levitated nanoparticles, which experience a multimode radiation pressure backaction (the photon recoil force) due to radiative scattering of the probe field. Here we extend the analysis of multimode backaction to compliant surfaces, accessing a broad class of mechanical resonators and fruitful analogies to quantum imaging. As an example, we consider imaging of the flexural modes of a membrane by sorting the spatial modes of a laser reflected from its surface. }
We show that backaction in this setting can be understood to arise from \textcolor{black}{spatiotemporal} photon shot noise, an effect that cannot be observed in single-mode optomechanics.  We also derive the imprecision-backaction product in the limit of purely spatial (intermodal) coupling, revealing it to be equivalent to the standard quantum limit for single-mode optomechanical coupling. Finally, we show that optomechanical correlations due to \textcolor{black}{spatiotemporal} backaction can give rise to two-mode entangled light, \textcolor{black}{providing a mechanism for entangling desired pairs of spatial modes}. In conjunction with high-$Q$ nanomechanics, our findings point to new opportunities at the interface of quantum imaging and optomechanics, including sensors and networks enhanced by spatial mode entanglement.


\end{abstract}

\maketitle
Active imaging protocols are ubiquitous in science and technology, and play a key role in the development of quantum optics \cite{wootters1979complementarity}. \textcolor{black}{Recently the quantum limits of active imaging have been explored in levitated optomechanics \cite{gonzalez2021levitodynamics}, with dielectric particles illuminated by strong free-space trapping fields.  In this setting, the standard quantum limit (SQL) arises due to interaction of the trapping field with vacuum fluctuations of electromagnetic modes into which the particle's motion is encoded, manifesting as both imprecision noise and a photon ``recoil force," which are inversely proportional.  In conjunction with efficient readout, measurements of sub-wavelength particles strong enough to observe the recoil force \cite{jain2016direct} have enabled feedback-based ground state cooling \cite{tebbenjohanns2021quantum,magrini2021real} and ponderomotive squeezed light generation \cite{magrini2022squeezed,militaru2022}. Techniques have also been explored to optimally image particle motion \cite{tebbenjohannsOptimalPositionDetection2019,tebbenjohannsOptimalOrientationDetection2022} and, or, evade multimode backation, using spatial mode sorting techniques \cite{dinter2024three} and tailored dielectric environments \cite{gajewski2024backaction,weiser2025backaction}
}.

Looking forward, an outstanding challenge is to extend quantum-limited imaging to broader class of 2D mechanical resonators, including a new class of ultrahigh-$Q$ flexural mode resonators \cite{engelsen2024ultrahigh} which can be functionalized for sensing and transduction \cite{metcalfe2014applications,engelsen2024ultrahigh,heinrich2021quantum}, and have recently been probed using optical levers \cite{hao2024back,prattNanoscaleTorsionalDissipation2023,pluchar2024quantum,choi2024quantum,shin2024laser}.  
Aside from alleviating practical constraints posed by \textcolor{black}{optical cavities \cite{CavityOptomechanics}}, the prospect of backaction-limited 2D imaging  offers intriguing opportunities at the interface of quantum imaging \cite{lugiatoQuantumImaging2002,defienne2024advances} and optomechanics \cite{aspelmeyerCavityOptomechanics2014}, including the application and generation of spatially entangled light, using ponderomotive effects \cite{chenEntanglementPropagatingOptical2020,barzanjehStationaryEntangledRadiation2019}. It also provides a platform for revisiting the physical description of multimode backaction in terms of spatiotemporal shot noise \cite{kim1994quantum,samphire1995quantum,loudon2003theory}\textcolor{black}{---where photon flux is random in both space and time---}in both pixel (Fig. 1) and paraxial mode bases commonly used in quantum imaging transceiver models \cite{fabre2020modes}.

\textcolor{black}{In this Letter, we explore spatiotemporal backaction in active imaging of a 2D mechanical resonator, and show that a novel form of ponderomotive spatial mode entanglement is accessible in the strong backaction regime.  The basis for our discussion is}
\begin{figure}[t!]
\centering  
\includegraphics[width=1\columnwidth]{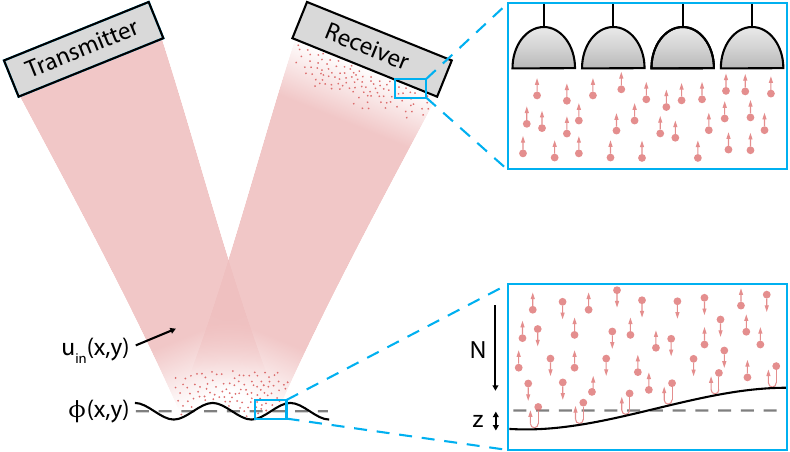}
\caption{\textcolor{black}{Pixel basis visualization of imprecision and backaction in active imaging of a vibrating membrane. Spatiotemporal shot noise at the receiver
gives rise to measurement imprecision.  The same noise produces radiation pressure backaction on the membrane.}}
\label{fig:Fig1} 
\vspace{-2mm}
\end{figure}
the thought experiment in Fig. \ref{fig:Fig1}, in which a vibrating membrane is illuminated by a laser beam (transmitter) and the reflected beam \cite{reflectivity} is imaged on a generic photoreceiver. Even if the intensity-averaged membrane displacement is zero\textcolor{black}{---corresponding to a vanishing static force---}spatially uncorrelated radiation pressure fluctuations yield a finite generalized \textcolor{black}{(modal)} force spectral density~\cite{pinardEffectiveMassQuantum,hao2024back,SI}
\begin{equation}
\label{eq:SFQBA1}
S_F^\t{BA} =8\hbar^2 k^2 \beta^2 N,
\end{equation}
where $N$ ($k$) is the laser photon flux (wavenumber) and
\begin{equation}
\label{eq:Beta1}
\beta^2 = \iint |u_\t{in}(x, y)|^2  \phi^2(x, y) dx dy,
\end{equation}
is a unitless overlap factor between the normalized transverse modeshape of the laser $u_\t{in}$ and the membrane $\phi$, respectively. 

For a generic receiver, it can furthermore be shown that 
\begin{equation}
\label{eq:3}
S_z^\t{imp} S_F^\t{BA} \ge \hbar^2,
\end{equation}
where $S_z^\t{imp}$ is the apparent membrane displacement $(z)$ spectral density due to \textcolor{black}{spatiotemporal} photon shot noise. 

The imprecision-backaction product in Eq. \ref{eq:3} corresponds to the SQL for continuous readout of a harmonic oscillator \cite{clerkIntroductionQuantumNoise2010}, and can be derived from a heuristic semi-classical model that assumes the photon flux is random in both space and time (see Appendix \cite{SI}).  Notably, it implies the existence of an optimal receiver \cite{OptimalReciever} even if the average phase shift of the reflected field is zero, corresponding to zero optomechanical coupling in conventional single-mode cavity optomechanics \cite{aspelmeyerCavityOptomechanics2014}.  A simple example is probing the angular displacement of the membrane near one of its vibrational nodes, for which the optimal receiver is one of a variety of beam displacement sensors (e.g., a lateral effect photodiode {\cite{fradgleyQuantumLimitsPositionsensitive2022}).  A non-trivial example is readout of a high order membrane vibration, which can be achieved using a spatial mode sorter \cite{boucherSpatialOpticalMode2020,choi2024quantum} or structured homodyne receiver \cite{sun2014smallQFI}, as discussed below.

To formally derive Eqs. \ref{eq:SFQBA1}-\ref{eq:3}, we now develop a Hamiltonian description of our thought experiment.  The essence of this approach, illustrated in Fig. \ref{fig2}, is to decompose the incident and reflected fields into orthogonal spatiotemporal modes, and to determine the energetic coupling between these modes, mediated by the membrane. 
Specializing to normal incidence, the spatial mode of the reflected field can be expressed as
\begin{subequations}\begin{align}
u_\t{in}e^{2ikz\phi}&\approx u_\t{in}+2i k z \beta u_\t{sc}\\
&=u_\t{in}+2ikz\left(\beta_\parallel u_\t{in}+\beta_\perp u_\perp\right)
\end{align}\end{subequations}
where $u_\t{sc} = \beta^{-1} u_\t{in}\phi$ is the spatial mode of the scattered field and 
\textcolor{black}{$u_\perp$ is its component orthogonal to $u_\t{in}$, normalized so 
\begin{equation}\label{eq:betaexpansion}
\beta^2 = \langle u_\t{in}\phi,u_\t{in}\phi \rangle = \beta_\perp^2 +\beta_\parallel^2    
\end{equation} 
where    $\beta_{\parallel(\perp)} = \langle u_{\t{in}(\perp)},u_\t{sc}\rangle$ and  
$\langle u, v\rangle  \equiv \iint u^* v dxdy$ is the spatial overlap integral \cite{uperp}.}
Assuming the incident and scattered fields are temporally orthogonal (confined to frequencies near the laser carrier $\omega_0$ and motional sidebands $\omega_0\pm\omega_\t{m}$, respectively), \textcolor{black}{the interaction Hamiltonian can be expressed as \cite{Hinteraction}}
\begin{subequations}\vspace{-4mm}\begin{align}
\hat{H}_{\t{int}} 
&= 2 \hbar k \beta (\hat{a}_\t{in} \hat{a}_{\t{sc}}^\dagger + \hat{a}_\t{in}^\dagger \hat{a}_{\t{sc}})  \hat{z} \!\\
& = 2 \hbar k \big[\beta_\parallel (\hat{a}_\t{in} \hat{a}^\dagger_\parallel + \hat{a}_\t{in}^\dagger \hat{a}_\parallel) \\
&\;\;\;\;\;\;\;\;\;+\beta_{\perp} (\hat{a}_\t{in} \hat{a}_\perp^\dagger + \hat{a}_\t{in}^\dagger \hat{a}_\perp) \big] \hat{z}
\end{align} \label{eq:Hinteraction}\end{subequations}
where $\hat{z}$ is the membrane displacement operator and $\hat{a}$ is the annihilation operator for each optical mode, normalized so that $\langle \hat{a}^\dagger \hat{a}\rangle $ is the photon flux  (see Appendix \cite{SI}).  For a laser in a strong coherent state, $\hat{a}_\t{in} \rightarrow \sqrt{N}+ \hat{a}_\t{in}$ \cite{RotatingFrame}, it follows that
\begin{equation}
\label{eq:Hint_linearized}
\hat{H}_{\t{int}} \approx 2 \hbar k \beta \sqrt{N} \hat{X}_{\t{sc}} \hat{z} \equiv \hat{F}_{\t{BA}} \hat{z},
\end{equation}
where $\hat{X}_{\t{sc}} = \hat{a}_\t{sc} + \hat{a}_\t{sc}^\dagger$ and $\hat{F}_{\t{BA}}$ are operators for the scattered field amplitude and radiation pressure force, respectively.

Equation \ref{eq:Hinteraction} and its linearized form, Eq. \ref{eq:Hint_linearized}, is similar to the canonical optomechanical  
Hamiltonian \cite{aspelmeyerCavityOptomechanics2014}, but is generalized to include the possibility of scattering into different spatial modes, $\beta_\perp \ne 0$ \textcolor{black}{(notably challenging to enhance with a cavity \cite{CavityOptomechanics})}.  A key insight from Eqs. \ref{eq:Hinteraction}-\ref{eq:Hint_linearized} is that radiation pressure backaction occurs due to mixing of the coherent laser field with vacuum fluctuations in the orthogonal interacting mode $u_\perp$. 
For purely dispersive, single mode optomechanical coupling \textcolor{black}{$\beta_\perp = 0$, the scattered field is only temporally orthogonal, and backaction can be understood as arising due to random photon arrivals in time.}  For purely spatial \textcolor{black}{(intermodal)} optomechanical coupling $\beta_\parallel =0$, radiation pressure backaction can be interpreted as arising from random photon arrivals in space. In the intermediate case ($\beta_\perp\beta_\parallel\ge 0)$, it is attributable to a combination of spatial and temporal shot noise. 

\begin{figure}
\centering
	\vspace{-2mm}
\includegraphics[width=0.8\columnwidth]{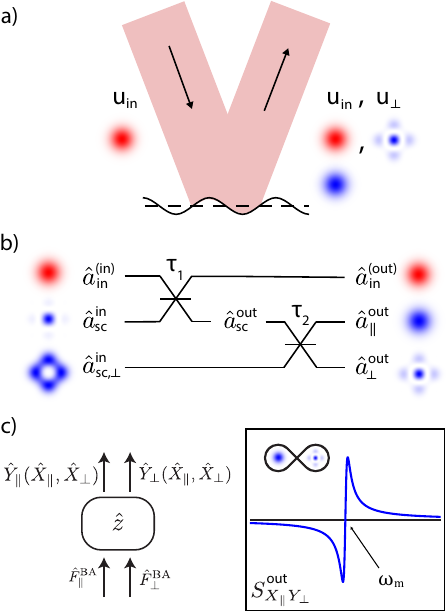}
\caption{\textcolor{black}{Paraxial mode} picture of spatiotemporal optomechanical coupling.  (a) Reflection of a laser from a vibrating surface \textcolor{black}{(here a fundamental Gaussian mode reflected from the node of a symmetric square membrane mode---see Fig. 4 for details)} yields three spatiotemporally orthogonal modes. Red and blue indicate frequency content near the laser carrier and motional sidebands, respectively.  (b) Beamsplitter model of the interaction, with splitting ratios $\tau_1(\omega) = 2 k\beta z(\omega)$ and $\tau_2 = \beta_\parallel/\beta_\perp$. $\hat{a}_\t{sc,\perp}$ represents sidebands in the orthogonal complement of $u_\t{sc}$. (c) Left: Schematic of optomechanical entanglement. $\hat{F}^\t{BA}_{\parallel(\perp)} = 2\hbar k\beta_{\parallel(\perp)}\hat{X}^\t{in}_{\parallel(\perp)}$ is the temporal (spatial) component of the backaction force (Eq. \ref{eq:Hint_linearized}). Right: Correlation spectrum between quadratures of the entangled modes (Eq. \ref{eq:twomodeentanglement}).}
\label{fig2}
 \vspace{-2mm}
\end{figure}

Equation \ref{eq:Hint_linearized} also gives insight into optimal (saturating Eq.~\ref{eq:SFQBA1} \cite{OptimalReciever}) receiver architectures, several of which are shown in Fig.~\ref{fig3}.  Specifically, for coherent illumination, displacement is encoded into the phase of the scattered field, as can be seen by computing the input-output relation (see Appendix \cite{SI})
\begin{equation}\label{eq:EOM}
\hat{a}_\t{sc}^\t{out}(t)  = \hat{a}_\t{sc}^\t{in} (t) + 2i\hbar k \beta \sqrt{N} \hat{z}(t).
\end{equation}
where $\hat{a}_\t{sc}^\t{out}$ ($\hat{a}_\t{sc}^\t{in}$) represent the field after (before) reflection. A known optimal receiver in this context is a ``structured" homodyne interferometer \cite{sun2014smallQFI,tebbenjohannsOptimalOrientationDetection2022} with a local oscillator mode $u_\t{LO} = u_\t{sc}$ (Fig. \ref{fig3}a).  \textcolor{black}{For $\beta_\parallel = 0$, another option is a} weighted pixel camera preceded by an optical Fourier transform  (Fig. \ref{fig3}b) \cite{pixelarray}.  Yet another is a spatial mode demultiplexer (SPADE) configured to distill $u_\t{sc}$ \cite{boucherSpatialOpticalMode2020,fontaine2019laguerre,ozer2022reconfigurable}, followed by homodyne detection. \textcolor{black}{(We note that both SPADE \cite{choi2024quantum,dinter2024three} and camera-based \cite{tavernarakis2025wavefront,werneck2024cross,minowa2022imaging} optomechanics have recently been explored.)}

\begin{figure}[t!]
\centering\includegraphics[width=0.93\columnwidth]{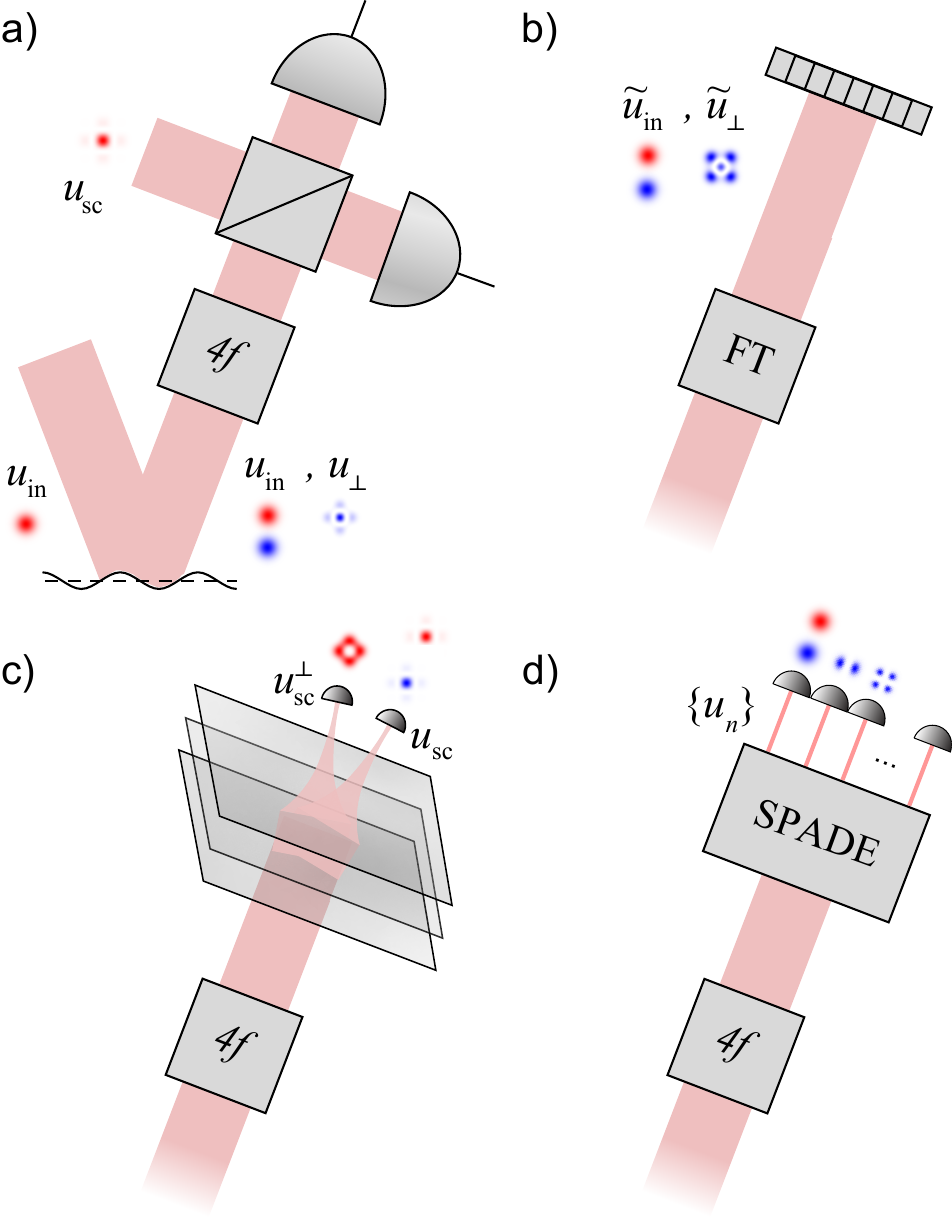}
\caption{\textcolor{black}{Receivers capable of optimal displacement readout \cite{OptimalReciever} for multimode optomechanical coupling $u_\t{sc}\ne u_\t{in}$.} (a) A ``structured" homodyne interferometer with its local oscillator in the scattered mode (a $4f$ system compensates for diffraction).  (b) An appropriately weighted pixel array \textcolor{black}{preceded by an optical Fourier transform, FT (e.g., by placing it in the far field)  \cite{SI}.} 
(c) A reconfigurable spatial mode sorter (SPADE) designed to distill the scattered mode. (d) A static SPADE sorting in an orthogonal (e.g., HG) basis.   }
\label{fig3}
\vspace{-2mm}
\end{figure}

Focusing on the structured homodyne receiver, which can access all quadratures of the scattered field, it is interesting to consider how imprecision,  backaction, and their correlation---ponderomotive squeezing in single-mode optomechanics---manifest in the case of multimode optomechanical coupling $u_\t{sc} \ne u_\t{in}$.  We thus consider the frequency-domain optomechanical equations of motion (see Appendix \cite{SI})
\begin{subequations}
\label{eq:outputQuads}
\begin{align}
\hat{X}_\t{sc}^\t{out}(\omega) &= \hat{X}_\t{sc}^\t{in}(\omega),\\
\hat{Y}_\t{sc}^\t{out}(\omega) &= \hat{Y}_\t{sc}^\t{in}(\omega)-4k\beta\sqrt{N}\hat{z}(\omega),\;\;\t{and}\\
\hat{z}(\omega) &= \chi(\omega)\left(F_\t{th}(\omega)+\hat{F}_\t{BA}(\omega)\right)\\
&= \chi(\omega)\left(F_\t{th}(\omega)-4\hbar k \beta \sqrt{N} \hat{X}_\t{sc}^\t{in}(\omega)\right),
\end{align}
\end{subequations}
where $\hat{Y}_\t{sc} = i(\hat{a}^\dagger_\t{sc} - \hat{a}_\t{sc})$ is the phase quadrature of the scattered mode, $\hat{X}_\t{sc}^\t{in}$ are $\hat{Y}_\t{sc}^\t{in}$ are input vacuum noise, $\chi$ is the mechanical susceptibility \cite{SI}, and $F_\t{th}$ is the thermal force. 

For generic quadrature $X_\t{sc}^\theta = X_\t{sc}\cos\theta+Y_\t{sc}\sin\theta$ with noise spectrum $S_{X_\t{sc}^\theta} = \cos^2\theta S_{X_\t{sc}} + \sin^2\theta S_{Y_\t{sc}} + \textcolor{black}{\sin(2 \theta)\t{Re} [S_{X_\t{sc}Y_\t{sc}}]}$ the apparent displacement spectral density $S_z^\theta$ is given by
\begin{equation}\label{eq:Sz_apparent}
S_z^\theta = \frac{S^\t{out}_{X_\t{sc}^{\theta}}}{16 k^2\beta^2 N \sin^2\theta} = S_z^\t{imp} + S_z^\t{BA} +S_z^\t{imp,BA} +S_z^\t{th}
\end{equation}
where (noting $S^\t{in}_{X_\t{sc}} = S^\t{in}_{Y_\t{sc}} = 2$) $S_z^\t{imp} = (8k^2\beta^2 N \sin^2\theta)^{-1}$ is the measurement imprecision due to photon shot noise, $S_z^\t{BA(th)}= |\chi|^2 S_F^\t{BA(th)}$ is the physical motion due to backaction (thermal) noise, and $S_z^\t{imp,BA}=(\textcolor{black}{8} k^2\beta^2 N \tan\theta)^{-1}\t{Re} [S^\t{out}_{X_\t{sc}Y_\t{sc}}] = - \textcolor{black}{2}\hbar\cot\theta\t{Re}[\chi]$ is the imprecision-backaction cross-spectrum \cite{SI}, corresponding to scattered mode quadrature correlations
\begin{equation}
S^\t{out}_{X_\t{sc}Y_\t{sc}} = - 16\hbar k^2\beta^2 N  \chi.
\end{equation}

Equation \ref{eq:Sz_apparent} is the general form for continuous linear displacement measurement of a mechanical oscillator \cite{clerkIntroductionQuantumNoise2010}, and yields the SQL in the limit of no correlations $S_z^\t{imp,BA} = 0$, corresponding to phase quadrature readout ($\theta = \pi/2$) and an imprecision-backaction saturating the lower bound of Eq. \ref{eq:3}.

In the case that imprecision-backaction correlations do not vanish $S_z^\t{imp,BA} \ne 0$, the physical meaning of Eq. \ref{eq:Sz_apparent} depends on the form of optomechanical coupling.  For purely dispersive coupling ($\beta_\perp = 0$), it corresponds to quadrature squeezing of the illumination field due to the effective optomechanical Kerr nonlinearity (ponderomotive squeezing \cite{aspelmeyerCavityOptomechanics2014}).  For purely spatial coupling ($\beta_\parallel = 0$), the same interpretation holds, but the mode of the squeezed state is orthogonal to that of the input field. 
For a mixture of dispersive and spatial optomechanical coupling, the optomechanical interaction gives rise to a two-mode entangled state, manifesting as non-zero correlations between the quantum fluctuations of the two orthogonal spatial modes comprising the scattered field (a basic requirement for demonstrating inseparability between continuous-variable \textcolor{black}{Gaussian} states  \cite{ duan2000inseparability, barzanjehStationaryEntangledRadiation2019, chenEntanglementPropagatingOptical2020,fabre2020modes}).  This can be seen by expressing Eq. \ref{eq:outputQuads} in terms of the $\{\hat{a}_\parallel,\;\hat{a}_\perp\}$ mode quadratures, $X_\t{sc}^\theta = \beta_\parallel X_\parallel^{\theta}+\beta_\perp X_\perp^{\theta}$, yielding (see Appendix \cite{SI})
\begin{equation}\label{eq:twomodeentanglement}   
S^\t{out}_{X_\parallel Y_\perp} = - 16 \hbar k^2 \beta_\perp\beta_\parallel N \chi,
\end{equation}
\textcolor{black}{
as illustrated in Fig. \ref{fig3}c. 
Notably, unlike previous demonstrations relying on single-mode cavity optomechanical coupling \cite{chenEntanglementPropagatingOptical2020,barzanjehStationaryEntangledRadiation2019}, this entanglement scheme requires only a single illumination field.  It also yields a strategy to selectively entangle two desired spatial modes ($u_\t{in},\;u_\perp$), by tailoring $\phi$ such that $u_\perp   = u_\t{in}(\phi - \beta_\parallel)/\beta_\perp$ \cite{HGLGentanglement}.   (In the limit $\beta_\parallel/\beta_\perp\ll 1$, for example, this amounts to $\phi \propto u_\perp/u_\t{in}$. )}

We now explore a concrete example illustrating the transition between dispersive and spatial optomechanical coupling.
As shown in Fig. \ref{fig4}, we consider a square membrane vibrating in a symmetric mode $\phi(x,y) = \cos(\pi x/\lambda_\t{m})\cos(\pi y/\lambda_\t{m})$ with nodal spacing $\lambda_\t{m}$.  The laser beam is taken to be in the fundamental Hermite-Gauss (HG) mode $u_{00}(x,y) =\sqrt{2/(\pi w_0^2)}e^{-((x-x_0)^2+(y-y_0)^2)/w_0^2}$ with diameter $2w_0 = 0.6\lambda_\t{m}$.  To visualize optomechanical coupling, we translate the beam position $(x_0, y_0)$ while monitoring the scattered mode expansion in the co-translated HG basis 
\begin{equation}
u_\t{sc} = \beta^{-1} \sum\langle u_{mn} ,\phi u_{00}\rangle u_{mn} \equiv \beta^{-1}\sum\beta_{mn}(x_0,y_0)u_{mn}
\end{equation}
noting that expansion coefficients $\beta_{mn}$ satisfy $\beta_\parallel = \beta_{00}$ and
\begin{equation}
\beta^2 = \sum|\beta_{mn}|^2.
\end{equation}

\begin{figure*}[ht]
\centering
\includegraphics[width=2\columnwidth]{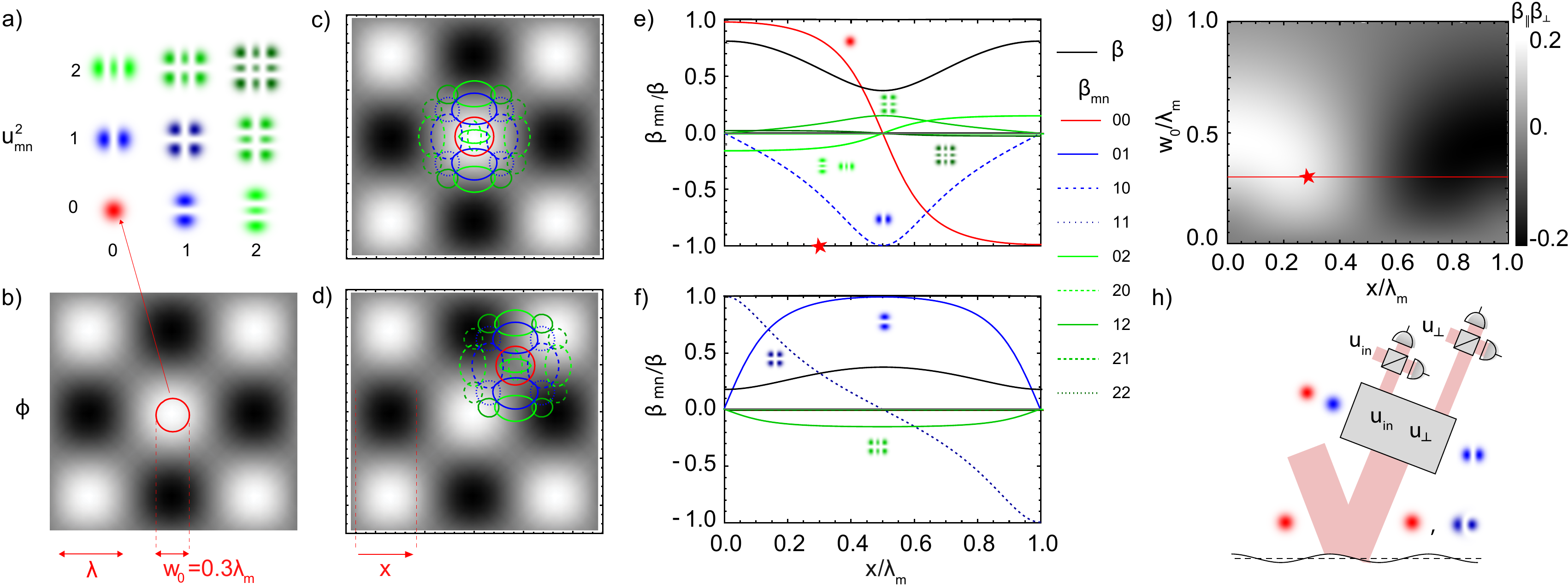}
\caption{Multimode optomechanical coupling to a membrane. (a) Hermite-Gauss (HG) mode intensity distributions $u^2_\t{mn}(x,y)$. (b) Mechanical mode shape $\phi(x,y)$ with nodal spacing $\lambda_\t{m}$.  The red circle represents the HG mode diameter $2w_0$. (c,d) Visualization of the overlap between $u_\t{mn}$ and $\phi$ when the beam is centered at an antinode (c) and node (d) of $\phi$. (e,f) Plots of the modal overlap factors $\beta_\t{mn}$ for $w_0 = 0.3\lambda_\t{m}$, versus lateral offset $x$ from the initial position in (c,d), respectively. (g) Density plot of \smash{$\beta_\perp\beta_\parallel$} versus mode waist $w_0$ and location $x$ as in (c,e), encoding the magnitude of the correlations between \smash{$Y_\perp^\t{out}$ and $X_\parallel^\t{out}$} as described in Eq. \ref{eq:twomodeentanglement}. (h) Dual-homodyne receiver for characterizing correlations between parallel $u_\t{in}$  and perpendicular $u_\perp$ scattered mode quadratures, e.g. $u_\t{in} = u_{00}$ and $u_\perp \approx u_{10}$ for the starred position in (e) and (g).}
\label{fig4}
\vspace{-2mm}
\end{figure*}

As shown in Fig. \ref{fig4}(e), translating the laser beam between an antinode and a node of the membrane mode gives access to different scattered modes and optomechanical couplings.  The case where the beam is centered on an antinode has been widely studied in the field of optomechanics and corresponds to mainly dispersive optomechanical coupling, $\beta_{\perp} \ll \beta_\parallel$.  The case where the beam is centered on a node has been little explored, and corresponds to purely spatial optomechanical coupling $\beta_\parallel = 0$.  In both cases, the magnitude of the overall coupling factor $\beta$ depends on the beam diameter.

We first emphasize the case where the laser beam is positioned halfway between antinodes ($x = \lambda_\t{m}/2$), so that the scattered mode is approximately HG$_{10}$ ($\beta_{10}\approx 1 $). In the limit $w_0\ll \lambda_\t{m}$ ($\beta_{10}\rightarrow 1)$, the membrane's motion can be modeled as an angular displacement $\varphi = 2\pi z/\lambda_\t{m}$ and backaction can be treated as a torque $\tau_\t{BA} = F_\t{BA}\lambda_\t{m}/2 $, satisfying 
\begin{equation}
S_{\varphi}^{\textcolor{black}{\t{imp}}} S_\tau^{\textcolor{black}{\t{BA}}} \ge \hbar^2.
\end{equation}
\textcolor{black}{This angular displacement measurement quantum} limit as has been recently studied using optical lever measurements performed on high-$Q$ nanomechanical resonators \cite{hao2024back,prattNanoscaleTorsionalDissipation2023, pluchar2024quantum, shin2024laser, choi2024quantum}.

As shown in Fig. \ref{fig4}e, translating the beam away from a node gives access to the more general case $\beta_\perp\beta_\parallel \ne 0$, \textcolor{black}{enabling} two-mode entanglement according to Eq. \ref{eq:twomodeentanglement}. For the highlighted (starred) position, if $w_0\ll \lambda_\t{m}$, entanglement is distributed between $u_\t{in}=u_{00}$ and $u_{\perp}\approx u_{10}$. This can be witnessed by performing a variational measurement \cite{kimble2001} in which the $u_\t{in}$ and $u_\perp$ ports of a reconfigurable mode sorter are analyzed using independent homodyne interferometers,  yielding measurements of \smash{$X_\perp^{\theta,\t{out}}$ and $X_\parallel^{\theta,\t{out}}$}, respectively.  In Fig. \ref{fig4}g we plot $\beta_\parallel \beta_\perp$ versus $x$ and $w_0$, illustrating that correlations are maximized at positions between a node and antinode.

In summary, we have explored radiation pressure back-action in the context of active imaging of a \textcolor{black}{2D} mechanical resonator, presenting a Hamiltonian description that includes two forms of optomechanical coupling.  In the first form, the incident and scattered field are in the same spatial mode ($\beta_\perp = 0$), corresponding to traditional dispersive optomechanical coupling. In the second form, the incident and scattered modes are orthogonal ($\beta_\parallel = 0$), corresponding to purely spatial optomechanical coupling.  Backaction in both cases originates from mixing of the incident field with vacuum fluctuations of the scattered mode, giving rise to a radiation pressure force which is random in time (for dispersive coupling) and space (for spatial coupling).  Recoil of the \textcolor{black}{resonator} gives rise to amplitude-phase correlations.  For purely dispersive or spatial coupling, these correlations correspond to single-mode ponderomotive squeezing. In the intermediate case $\beta_\parallel\beta_\perp\ne 0$, they correspond to generation of two-mode entanglement.

Looking forward, advances in free-space optomechanics augur well for the study of spatiotemporal backaction, including recent demonstrations of ponderomotive squeezing with levitated nanospheres \cite{militaru2022, magrini2022squeezed} and high cooperativity deflectometry of nanobeams \cite{hao2024back,pluchar2024quantum,shin2024laser}. 
To extend these experiments, insights might be drawn from a growing portfolio of ``quantum-inspired" imaging protocols, such as super-resolution imaging \cite{grace2022Hypothesistesting,Tsang2016quantum} with low-loss SPADE technology \cite{boucherSpatialOpticalMode2020}.  Entanglement-enhanced imaging is also naturally applicable to mechanical resonators, and provides a complementary approach to entanglement-enhanced distributed force sensing \cite{xia2023entanglement}. 
Finally, it is interesting to consider nanomechanical resonators as quantum imaging testbeds, since they provide a diversity of multi-parameter estimation problems, and a rubric for transceiver design, through the backaction-imprecision product (Eq. \ref{eq:3}) \cite{tsangFundamentalQuantumLimit2011}. \textcolor{black}{An important example is simultaneously estimating the displacement of multiple modes, whose applications include impulse sensing \cite{schmerling2025optimal}, mass spectroscopy \cite{sader2024data}, photothermal deflectometry \cite{west2023photothermal}, and force microscopy \cite{nievergelt2014high}.}
Combining these considerations may usher in a new era of imaging-based quantum optomechanics.

\stoptoc
\section*{Acknowledgments}


The authors thank Ewan Wright, Allison Rubenok, and Morgan Choi for helpful discussions.  This work was supported by the National Science Foundation (NSF) through award No. 2239735. CMP acknowledges support from the ARCS Foundation. WH and SG acknowledge support from Air Force Office of Scientific Research contract No. FA9550-22-1-0180. ND acknowledges support from the NSF Graduate Research Fellowship under Grant No. DGE-2137419.
\bibliography{ref}

\begin{thebibliography}{64}
\bibitem [{ref()}]{reflectivity}%
  \BibitemOpen
  \href@noop {} {}\bibinfo {note} {For simplicity, we assume a perfect membrane
  reflectivity, $R = 1$. In general, $S_F^\t{BA}\propto R$ and
  $S_z^\t{imp}\propto R^{-1}$, so Eq. 3 is preserved.}\BibitemShut {Stop}%
\bibitem [{upe()}]{uperp}%
  \BibitemOpen
  \href@noop {} {}\bibinfo {note} {$\{u_\t{in},u_\perp\}$ forms an orthonormal
  basis for all $u_\t{sc}$, such that $\langle u_\t{sc},u_\t{sc} \rangle =
  \beta^2 = \beta_\perp^2 + \beta_\parallel^2$, where $\beta_{\parallel(\perp)}
  = \langle u_\t{sc},u_{\t{in}(\perp)}\rangle$.}\BibitemShut {Stop}%
\bibitem [{Hin()}]{Hinteraction}%
  \BibitemOpen
  \href@noop {} {}\bibinfo {note} {Eq. 5 is the part of the total Hamiltonian
  which includes the optomechanical interaction, and is time dependent, since
  carrier depletion is ignored. As discussed in the Appendix, $a_\text{in}(t)$
  and $a_\text{sc}(t)$ are also time-dependent Fourier-transformed
  operators.}\BibitemShut {Stop}%
\bibitem [{Rot()}]{RotatingFrame}%
  \BibitemOpen
  \href@noop {} {}\bibinfo {note} {Here and below, we define operators in a
  frame rotating at the laser carrier frequency, $\omega_0$.}\BibitemShut
  {Stop}%
\bibitem [{pix()}]{pixelarray}%
  \BibitemOpen
  \href@noop {} {}\bibinfo {note} {\textcolor{black}{Direct imaging is optimal
  under certain conditions, as described in the appendix. For the
  Fourier-transform imaging scheme in Fig. 3b, it is necessary that
  $\beta_\parallel = 0$ and
  $\text{Arg}[\tilde{u}_\text{in}/\tilde{u}_\text{sc}]=\pi/2$.}}\BibitemShut
  {Stop}%
\bibitem [{HGL()}]{HGLGentanglement}%
  \BibitemOpen
  \href@noop {} {}\bibinfo {note} {Note that because $\phi$ cannot impart a
  helical phase, $u_\t{in}$ and $u_\perp$ must have the same orbital angular
  momentum.}\BibitemShut {Stop}%
\end{thebibliography}
\resumetoc

\AtEndDocument{\stoptoc
\section*{Supplementary Information for ``Imaging-based Quantum Optomechanics"}
Below we provide detailed derivations of Eqs. 1-12 in the main text.  Section III provides semiclassical derivations of spatiotemporal radiation pressure quantum backaction and imprecision noise for several receiver architectures. Section IV derives a multi-mode optomechanical Hamiltonian in the paraxial approximation and simplifies to a two-mode interaction Hamiltonian in the case of single mode illumination.  Section V derives the input-output relations for interacting modes for the case of single-mode illumination. 

\resumetoc

\tableofcontents

\color{black}

\section{Definitions}

For time-dependent variable $X(t)$ we define the Fourier transform
\begin{equation}\label{eq:FT}
    X(\omega) = \int_{-\infty}^{\infty} X(t)e^{i\omega t} dt,
\end{equation}
and the single-sided power spectral density (PSD)
\begin{equation}\label{eq:SSPSD}
S_{X}(\omega>0) = 2\int_{-\infty}^{\infty} \langle \{X(t),X(t+\tau)\}\rangle e^{i\omega \tau} d\tau,
\end{equation}
where $\{a,b\}$ $=(ab+ba)/2$ is the symmetrization operator and $\langle,\rangle$ denotes the time average.  The PSD is normalized such that $\langle X^2\rangle = \int_0^\infty S_X(\omega) d\omega/(2\pi)$.

For two variables $X_1(t)$ and $X_2(t)$, we likewise define the single-sided cross spectral density (CSD)
\begin{equation}\label{eq:SSCSD}
S_{X_1 X_2}(\omega>0) = 2\int_{-\infty}^{\infty} \langle \{X_1(t),X_2(t+\tau)\}\rangle e^{i\omega \tau} d\tau.
\end{equation}
We occasionally use the notation $S_{X_1,X_2}$ and $S_X^{1,2}$ instead.\\

For tranverse modeshapes $f(x,y)$ and $g(x,y)$, we define the spatial overlap integral
\begin{equation}
\langle f, g\rangle \equiv \iint_{-\infty}^\infty f^*(x,y)g(x,y)dxdy.
\end{equation}

For convenience, all optical $u$ and mechanical $\phi$ modeshapes in the main text and below are normalized such that 
\begin{equation}
    \langle u, u\rangle = \langle \phi, \phi\rangle = 1.
\end{equation}
From this convention, it follows that if $\langle u_i,u_j\rangle = \delta_{ij}$ and $u\phi = \sum_{i} \langle u_i,u\phi\rangle u_i$, then $\langle u\phi, u\phi\rangle = \sum_{i}\langle u_i,u\phi\rangle ^2$. 


\onecolumngrid

\section{Table of variables}

Below is a list of the variables used in the main text and in this appendix, in approximate order of appearance. 

\begin{longtable}{ |p{2.7cm}||p{\textwidth-3.5cm}|  }
 \hline
 \multicolumn{2}{|c|}{Introduced in the main text} \\
 \hline
 $k, \omega_0, N, u_\t{in} $ & Wavenumber $(k)$, angular frequency ($\omega_0)$, photon flux $(N)$, and transverse modeshape $(u_\t{in})$ of incident laser\\
 \hline
  $z$ \textcolor{black}{($z_0$)}, $\phi$ & Displacement ($z$, \textcolor{black}{denoted $z_0$ in the appendix}) and modeshape
 $(\phi)$ of mechanical oscillator\\
   \hline
   \rule{0pt}{2.4ex}$\beta$ & Intensity overlap of mechanical and optical modeshape, $\beta = \sqrt{\langle u_\t{in}\phi,u_\t{in}\phi\rangle} $\\
 \hline
     \rule{0pt}{3ex}$\beta_\parallel$, $\beta_\perp$ & Temporal ($\beta_\parallel = \langle u_\t{sc},u_\t{in}\rangle $) and spatial  ($\beta_\perp = \langle u_\t{sc},u_\perp\rangle $) fraction of $\beta = \sqrt{\beta_\parallel^2+\beta_\perp^2}$ \\
 \hline
  $F_\t{BA}$, $F_\t{th}$  & Backaction force ($F_\t{BA}$) and thermal force  ($F_\t{th}$)\\
  \hline
   $u_\t{sc}$, $u_\perp$ & Modeshape of scattered field ($u_\t{sc}$) and interacting mode ($u_\perp$), satisfying $\beta u_\t{sc} = \phi u_\t{in} = \beta_\parallel u_\t{in}+\beta_\perp 
 u_\perp$ \\
 \hline
    \rule{0pt}{2.5ex} 
 $\hat{H}, \hat{H}_\t{int}, \hat{H}_\t{m}, \hat{H}_\t{EM}$ & Total ($\hat{H}$), interaction ($\hat{H}_\t{int}$), mechanical ($\hat{H_\t{m}}$), and electromagnetic field ($\hat{H}^\t{free}_\t{EM}$) Hamiltonian\\
 \hline
$\hat{a}^{(\dagger)}_\t{in},\hat{a}^{(\dagger)}_\t{sc},\hat{a}^{(\dagger)}_\perp,  \hat{a}_\t{sc}^{(\dagger) \t{out}}$ & Annihilation (creation) operator for input, scattered, interacting, and output modes\\
 \hline
 $\hat{X}_\t{in}(\hat{Y}_\t{in}),X_\t{sc}(\hat{Y}_\t{sc})$ & Amplitude (phase) quadrature operator for input and scattered modes\\
 \hline
     \rule{0pt}{2.3ex} $\hat{X}^\theta$,$\theta$ & Rotated (by angle $\theta$) quadrature operator, $\hat{X}^\theta = \hat{X}\cos(\theta)+\hat{X}\sin(\theta)$\\
 \hline
 $\omega_\t{m}, \gamma_\t{m}, \chi, m_\t{eff}, z_\t{zp}$ & Mechanical resonance frequency ($\omega_\t{m}$), damping rate ($\gamma_\t{m}$), susceptibility ($\chi$), effective mass ($m_\t{eff}$), and zero-point displacement ($z_\t{zp}$). \\
 \hline
 $S_{X_\t{sc}Y_\t{sc}}, S^\t{out}_{X_\parallel Y_\perp},  S_z^\t{BA, imp}$ & CSD of the output scattered mode amplitude and phase ($S_{X_\t{sc}Y_\t{sc}}$), output temporal and spatial modes ($S^\t{out}_{X_\parallel Y_\perp}$), and backaction-imprecision noise ($S_z^\t{BA, imp}$) \\
 \hline
 \rule{0pt}{2.5ex}    
$S_z^{\t{BA}}, S_z^\t{th}, \newline S_z^\t{imp} (S_\phi^\t{imp}),  S_z^\theta$ & Backaction ($S_z^{\t{BA}}$), thermal ($S_z^{\t{th}}$), (angular displacement) imprecision ($S_z^{\t{imp}}$, $S_\phi^{\t{imp}}$), and apparent displacement ($S_z^\theta$) PSD \\
 \hline
 $\lambda_\t{m}, \varphi$ & Membrane antinode spacing ($\lambda_\t{m}$) and angular displacement ($\varphi$) \\
 \hline
 $\tau_\t{BA}, S_\tau^\t{BA}$ & Backaction torque ($\tau_\t{BA}$) and backaction torque PSD ($S_\tau^\t{BA}$)\\
 \hline
 $u_{\textcolor{black}{mn}} (w_0), U_{kn}$ & $n, m^\t{th}$ Hermite Gauss mode ($u_{\textcolor{black}{mn}}$, with beam waist $w_0$), general paraxial beam mode with wavenumber $k$ and mode number $n$ ($U_{kn}$)\\
 \hline
  \multicolumn{2}{|c|}{Introduced in the SI}\\
 \hline
 $I_\t{in}, \delta I_\t{in}^\t{shot}$ & Photon intensity ($I_\t{in}$) and shot noise ($\delta I_\t{in}^\t{shot}$) \\
 \hline
 $S_{I,I'}^\t{in,shot}, S_N$ & Photon intensity ($S_{I,I'}^\t{in,shot}$) and photon number ($S_N$) shot noise PSD\\
 \hline
 $\delta P_\t{rad}^\t{shot} (S_\t{P, P'}^\t{rad,shot})$ & Radiation pressure shot noise ($\delta P_\t{rad}^\t{shot}$) and radiation pressure shot noise (CSD)\\
 \hline
 $E_\t{LO (ref, in)}, \newline A_\t{LO}, u_\t{LO, (ref, in)}, \newline N_\t{LO}, \theta_\t{LO (in)}$ & Local oscillator (reflected, input) electric field \textcolor{black}{$(E_\t{LO (ref, in)})$}, electric field amplitude \textcolor{black}{$(A_\t{LO})$}, modeshape \textcolor{black}{$u_\t{LO, (ref, in)}$}, photon flux \textcolor{black}{($N_\t{LO}$)}, and phase \textcolor{black}{($\theta_\t{LO (in)}$)} \\
 \hline
 $i_\t{hom}, i_{\textcolor{black}{mn}}$ & Homodyne ($i_\t{hom}$) and pixel photocurrent ($i_{\textcolor{black}{mn}}$) \\
 \hline
 $\delta i^\t{shot}_{\textcolor{black}{mn}}, \sigma^2_{\delta i_{\textcolor{black}{mn}}}, S_{\delta i_{\textcolor{black}{mn}}}, \Delta_f$ & Pixel shot noise ($\delta i^\t{shot}_{\textcolor{black}{mn}}$), pixel shot noise variance ($\sigma^2_{\delta i_{\textcolor{black}{mn}}}$), and pixel shot noise PSD ($S_{\delta i_{\textcolor{black}{mn}}}$), in measurement bandwidth $\Delta_f$\\
 \hline 
 $\xi, d, l$ & Photodetection efficiency ($\xi$), membrane-camera separation ($d$) and pixel width ($l$)\\
 \hline
 $\textbf{E, B}, \textbf{e}, \textbf{b}$ & Electric ($\textbf{E}$) and magnetic ($\textbf{B})$ field, electric ($\textbf{e}$) and magnetic ($\textbf{b}$) field unit vector \\
 \hline
 $\epsilon_0, \mu_0, \eta$& Permittivity ($\epsilon_0)$ and permeability ($\mu_0$) of free space and electromagnetic impedance ($\eta)$\\
 \hline
 \color{black}$\beta_{n,n'}$ ($\beta_{mn}$) &  \color{black}Transverse spatial overlap $\langle u_n|\phi|u_{n'}\rangle $ between optical modes $u_n$ and $u_n'$. (In the main text, for notational simplicity, we define $\beta_{mn}  \equiv \beta_{mn,00} = \langle u_{mn}|\phi|u_{00}\rangle$ between HG$_{mn}$ and HG$_{00}$ modes.)\color{black} \\

 \hline
 $\hat{X}_+, \hat{Y}_{-}, I(\omega)$ & Joint amplitude ($\hat{X}_+)$ and phase ($\hat{Y}_-$) quadrature of temporal and spatial output modes, Duan-Giedke-Cirac-Zoller inseparability criterion ($I(\omega)$)\\
 \hline

\end{longtable}

 \vspace{1mm}
\hrulefill
\vspace{7mm}

\twocolumngrid


\justifying
\color{black}

\vspace{-2mm}
\section{Spatiotemporal Shot Noise:\\ Semiclassical Model}
\vspace{-2mm}

Consider a laser beam with mean photon flux $N$ (units $\t{s}^{-1}$) focused onto a surface in the $x-y$ plane.  The photon intensity on the surface (units $\t{s}^{-1} \t{m}^{-2}$) can be expressed as  
\begin{equation}
	I_\t{in}(x,y) = N \left|u_\t{in}(x,y)\right|^2
\end{equation}
where $u_\t{in}(x,y)$ is the transverse beam modeshape with normalization $\iint dx dy \left|u_\t{in}(x,y)\right|^2 =1$ .

Semiclassically, photon shot noise can be modeled as a spatiotemporally random intensity fluctuation $\delta I_\t{in}^\t{shot}(t,x,y)$ with CSD
\begin{equation} \label{eq:intensityCrossPSD}
	S_{I,I'}^\t{in,shot} = 2 I_\t{in}(x,y) \delta(x-x')\delta(y-y')\textcolor{black}{.}
\end{equation}
As a consistency check, note that the mean photon flux through a subsurface $A$ is $N_A = \iint_A dxdy \, I_\t{in}(x,y)$ and that shot noise contributes temporal fluctuations $\delta N_A^\t{shot}(t) = \iint_A dxdy \, \delta I_\t{in}^\t{shot}(t,x,y)$ through this subsurface with PSD
\begin{equation}\label{eq:Nshotnoise}
	S_{N_A }^\t{shot} = \iint_{A } dxdy \iint_{A } dx'dy' \, S_{I, I'}^\t{in,shot}  = 2N_A, 
\end{equation}
which is the standard result for temporal shot noise~\cite{clerkIntroductionQuantumNoise2010}.

\vspace{-2mm}
\subsection{Generalized Radiation Pressure Backaction}
\vspace{-2mm}

We now consider the influence of shot noise on active imaging of a compliant landscape, starting with measurement backaction.  Towards this end, following Fig. 1, suppose the laser is focused onto a perfectly reflective membrane vibrating in a transverse mode with amplitude $z_\t{m}(x,y,t)= z_0(t)\phi(x,y)$. Photon shot noise imparts a spatiotemporally random radiation pressure $\delta P^\t{shot}_\t{rad}(t,x,y)
= 2\hbar k \delta I_\t{in}(t,x,y)$ on the membrane with CSD $S_{P,P'}^\t{rad,shot} = 4 \hbar^2 k^2 S_{I,I'}^\t{in,shot}$, where $k$ is the laser wavenumber.  The resulting generalized backaction force on the membrane mode is given by
\begin{equation}
	F_\t{BA}(t) = \iint dxdy \, \phi(x,y) \delta P_\t{rad}^\t{shot}(t,x,y).
\end{equation}
The associated force PSD is given by
\begin{subequations}\label{BAderived}\begin{align} 
		S_{F }^\t{BA} &= \iiiint dx dy dx'dy' \phi(x,y)\phi(x',y')\, S_{P, P'}^\t{rad,shot} \\
		& = 8 \hbar^2 k^2 \beta^2 N
\end{align}\end{subequations}
with
\begin{equation}\label{eq:betaSI}
	\beta^2 \equiv  \iint dx dy \, \phi^2(x,y)\left|u_\t{in}(x,y) \right|^2.
\end{equation}

\subsection{Imprecision Noise and Receiver Ideality}

Shot noise also gives rise to imprecision in an imaging-based estimate of the membrane displacement $z_0$, with a magnitude that depends on the receiver architecture.  The receivers in Fig. 3 all infer $z_0$ from the complex amplitude of the reflected field, given for small displacement ($k z_0\ll 1$) as
\begin{subequations}\begin{align}
		E_\t{ref}(x,y) &=  E_\t{in}(x,y) e^{i 2 kz_0 \phi(x,y)}\\
		&\approx A_\t{in}\left( u_\t{in}(x,y) + i2\beta k z_0 u_\t{sc}(x,y) \right)
\end{align}\end{subequations}
where $A_\t{in} = \sqrt{2\eta \hbar k c N}e^{i\theta_\t{in}}$ is the incident field amplitude, $\eta$ is the electromagnetic impedance, and $u_\t{sc}(x,y)\equiv u_\t{in}(x,y)\phi(x,y)/\beta$ is the modeshape of the scattered field. Below, we explore two examples: a structured homodyne interferometer (Fig. 3a) and a pixelated camera in the far field (Fig. 3b). We show that both receivers are ideal in the sense that their displacement imprecision $z_\t{imp}$ due to shot noise saturates the Standard Quantum Limit (SQL): $S_{z}^\t{imp}  S_{F }^\t{BA} = \hbar^2$.

\vspace{-2mm}
\subsubsection{Structured Homodyne Interferometer}
\vspace{-2mm}

In a structured homodyne interferometer, the reflected field is combined on a 50:50 beamsplitter with a local oscillator field $E_\t{LO}(x,y) = A_\t{LO} u_\t{LO}(x,y)$ with amplitude $A_\t{LO} = e^{i\theta_\t{LO}}\sqrt{2\eta \hbar k c N_\t{LO}}$ and modeshape $u_\t{LO}$.  Balanced detection of the \textcolor{black}{beamsplitter} outputs yields a photocurrent
\begin{subequations}\begin{align}
		i_\t{hom} &= 4k z_0 \beta\xi\sqrt{N_\t{LO}N}\langle u_\t{LO},u_\t{sc}\rangle \sin(\theta_\t{LO}-\theta_\t{in})\\
		&\le 4 kz_0 \beta \sqrt{N_\t{LO}N}
\end{align}\end{subequations}
where $\xi$ is the photodetection efficiency. 
The upper bound is achieved when $\xi = 1$, $u_\t{LO} = u_\t{sc}$ and $\theta_\t{LO}-\theta_\t{in} = \pi/2$.

\textcolor{black}{The displacement imprecision $S_z^\t{imp}$ corresponds to the membrane displacement that would produce a photocurrent fluctuation commensurate with shot noise. $S_z^\t{imp}$ is determined by
dividing the photocurrent shot noise $S_i^\t{hom,shot} = 2\xi(N_\t{LO}+N)$ (with units amps/Hz) by the displacement-photocurrent tranduction factor $d i_\t{hom}/d z_0$ (with units amps/meter):} 
\begin{equation}
	S_z^\t{imp} \equiv \left(\frac{d i_\t{hom}}{d z_0}\right)^{-2}S_i^\t{hom,shot}\ge \frac{1}{8 k^2 \beta^2 N},
\end{equation}
with the lower bound achieved with the above parameters and a strong local oscillator, $N_\t{LO}\gg N$. Comparison with Eq. \ref{BAderived} reveals this bound to saturate the SQL: $S_{z}^\t{imp}  = \hbar^2 / S_{F }^\t{BA}$.

\vspace{-2mm}
\subsubsection{Pixelated camera in the far field}
\vspace{-2mm}
Consider a pixelated camera imaging the reflected field in the far field as shown in Fig. 3b---i.e., at a distance $d\gg kw_0^2$ from the membrane, where $w_0$ is the laser spot size. Using Fraunhofer's diffraction formula 
and assuming $k z_0 \ll 1$, the photon intensity on the camera can be expressed as
\textcolor{black}{\begin{equation} \label{eq:cameraIntensity}
	I(x,y) \approx N \left(\left|\tilde{u}_\t{in} \right|^2 - 4\beta k z_0 \t{Im} \left\{ \tilde{u}_\t{sc}\tilde{u}^*_\t{in} \right\}\right)
\end{equation}
where $\tilde{u}(x,y) = (\lambda d)^{-2}\iint dx' dy' e^{-i k (xx'+yy')/d} u(x',y')$ is the normalized spatial Fourier transform of $u(x,y)$.}

Information about $z_0$ is encoded in the photocurrent produced by each pixel, $i_{mn}$.  Assuming small, square pixels with side length $l$ and coordinates $(x_m,y_n)$ in the detector plane, the photocurrent  mean and fluctuations can be approximated as
\color{black}\begin{subequations}\begin{align}
		\bar{i}_{mn} &= N\xi \left|\tilde{u}_{\t{in},mn}\right|^2 l^2 \;\;\t{and}\\
		\delta i_{mn}(t) &= -4\beta kz_0(t) N \xi B_{mn} l^2 + \delta i_{mn}^\t{shot}(t)
\end{align}\end{subequations}\color{black}
respectively, where subscript $mn$ denotes evaluation at position  $(x_m,y_n)$, $B_{mn}\equiv \t{Im} \left\{ \tilde{u}_{\t{sc},mn}\tilde{u}^*_{\t{in},mn} \right\}$, and $\delta i_{mn}^\t{shot}$ is the photocurrent shot noise with PSD $S_{\delta i_{mn}}^\t{shot} = 2\bar{i}_{mn}$. 

To make an estimate $z_0^\t{est}$ of $z_0$, the sum of squared residual photocurrents can be minimized, weighted by the shot noise inverse variance, viz.
\begin{equation}\label{eq:cameraleastsquares}
	\t{Min}\sum_{mn}\sigma_{i_{mn},\t{shot}}^{-2} \left( \frac{\partial i_{mn}}{\partial z_0}z_0^\t{est} - \delta i_{mn}\right)^2
\end{equation}
where 
\begin{equation}
	\sigma^2_{i_{mn},\t{shot}} = S_{\delta i_{mn}}^\t{shot} \Delta_f = 2 \bar{i}_{mn} \Delta_f
\end{equation}
is the variance of the pixel shot noise in bandwidth $\Delta_f$.

The solution to Eq. \ref{eq:cameraleastsquares}, $z_0^\t{est}=z_0  + z_\t{imp}$, has an imprecision
\begin{equation}
	z_\t{imp}= \left(\sum_{mn}\sigma_{i_{mn},\t{shot}}^{-2}\left(\frac{\partial i_{mn}}{\partial z_0}\right)^{2}\right)^{-1} \sum_{mn}\sigma_{i_{mn},\t{shot}}^{-2} \frac{\partial i_{mn}}{\partial z_0}\delta i_{mn}^{\t{shot}}
\end{equation}
To compute $S_z^\t{imp}$, we note that shot noise is uncorrelated between pixels 
and take the continuum limit $l\rightarrow 0$.  A straightforward calculation reveals $S_{z}^\t{imp} = \kappa \hbar^2/S_F^\t{BA}$, where \color{black}
\color{black}\begin{equation}\label{eq:pixelatedcameraideality}
	\begin{aligned}
		\kappa &\equiv  \left(\iint dxdy\, \frac{\left(\t{Im}\left[\tilde{u}_\t{sc}(x,y)\tilde{u}^*_\t{in}(x,y)\right]\right)^2}{\left|\tilde{u}_\t{in}(x,y)\right|^2} \right)^{-1} \\	
		& = \frac{\beta^2}{{\beta_\perp}^2} \left(\iint dxdy\, |\tilde{u}_\perp(x,y)|^2 \sin^2\left(\t{Arg}\left[\frac{\tilde{u}_\perp(x,y)}{\tilde{u}_\t{in}(x,y)}\right]\right) \right)^{-1} \\	
	\end{aligned}
\end{equation}\color{black}
is a unitless ideality factor and $u_\perp$ is the orthogonal component of $u_\t{sc}$ ($\beta u_\t{sc} = \beta_\perp u_\perp + \beta_\parallel u_\t{in}$) as defined in the main text.

Equation \ref{eq:pixelatedcameraideality} implies that the ideality of the far-field camera receiver depends on the relationship between the input $u_\t{in}$ and scattered $u_\t{sc}$ modeshapes---i.e., the nature of the optomechanical coupling. For example, the camera provides no displacement information ($\kappa = \infty$) in the case of purely dispersive coupling, $u_\t{sc} = u_\t{in}$.  It can however perform an ideal measurement ($\kappa = 1$) in the case of purely spatial optomechanical coupling $\beta = \beta_\perp$, \textcolor{black}{provided that the relative phase between $\tilde{u}_\perp$ and $\tilde{u}_\t{in}$ is $\pi/2$ (in essence, performing a homodyne phase measurement of the scattered field, using the incident field as a local oscillator).  This situation arises, for example, when the incident and scattered fields are in Hermite-Gauss modes $u_{mn}$ of opposite total parity $|m+n-m'-n'| =2\mathbb{Z}^++1$.  The canonical example is an optical lever, in which a laser in the fundamental Guassian mode $u_{00}$ is reflected off a rotating surface ($\phi \propto x$), producing a scattered field in mode $u_\t{sc} = u_{01}$.} 

\vspace{-2mm}
\section{Hamiltonian Treatment}
\vspace{-2mm}

We now derive the effective \textcolor{black}{Hamiltonian} for the optomechanical interaction discussed in the main text (Eqs. 5-6).  Following the above treatment, we consider a perfectly reflective membrane located at $z=0$, vibrating in a transverse mode with amplitude $z_\t{m}(x,y,t)= z_0(t)\phi(x,y)$.  We assume a monochromatic laser beam is incident on the membrane from the left ($z<0$). The electromagnetic field $\{\textbf{E},\textbf{B}\}$ in this domain is a closed system with classical Hamiltonian,
\begin{equation}\label{eq:HEM_L}
	H_\t{EM} = \int_{-\infty}^\infty\int_{-\infty}^\infty dxdy\int_{-\infty}^{z_\t{m}} \left(\frac{\epsilon_0\textbf{E}(\textbf{r})^2}{2} + \frac{\textbf{B}(\textbf{r})^2}{2\mu_0}\right)dz,
\end{equation}
where $\epsilon_0$ ($\mu_0$) is the permittivity (permeability) of free space. For sufficiently small $z_\t{m}$, we can approximate
\begin{subequations}\begin{align}\label{eq:HEM_L_free}
		&H_\t{EM} \approx \iint dxdy\int_{-\infty}^{0} \left(\frac{\epsilon_0\textbf{E}(\textbf{r})^2}{2} + \frac{\textbf{B}(\textbf{r})^2}{2\mu_0}\right)dz\\\label{eq:HEM_L_int}
		&+z_0\iint dxdy\left(\frac{\epsilon_0\textbf{E}(x,y,0)^2}{2} + \frac{\textbf{B}(x,y,0)^2}{2\mu_0}\right) \phi(x,y)\\
		&\;\;\;\;\;\;\;\;\equiv H_\t{EM}^\t{free}+ H_\t{int},
\end{align}\end{subequations}
where $H_\t{EM}^\t{free}$ (Eq. \ref{eq:HEM_L_free}) describes the free field evolution and $H_\t{int}$ (Eq. \ref{eq:HEM_L_int}) describes the optomechanical interaction.  

We now wish to quantize $H_\t{EM}$. To this end, following an standard approach \cite{kim1994quantum,loudon2003theory,samphire1995quantum}, we quantize the paraxial fields in the half-space $z<0$ \cite{loudon2003theory}, assuming normal incidence and the boundary condition $E(x,y,0)=0$ for a perfect reflector:
\begin{subequations}\begin{align}
		\hat{\textbf{E}}(\textbf{r},t) 
		& = -\textbf{e}\sum_{n}\int_0^\infty \t{d}\omega \sqrt{\frac{\hbar \omega}{ \pi \epsilon_0 c}} \Large(\hat{a}_{n}(\omega)U_{kn}(\textbf{r})e^{-i\omega t} \\
        &\hspace{40mm}+\t{h.c.}\Large)\sin(kz)\notag\\
		\hat{\textbf{B}}(\textbf{r},t) &= \frac{i}{c}\textbf{b}\sum_{n}\int_0^\infty \t{d}\omega \sqrt{\frac{\hbar \omega}{ \pi \epsilon_0 c}} \Large(\hat{a}_{n}(\omega)U_{kn}(\textbf{r})e^{-i\omega t}\\&\hspace{40mm} -\t{h.c.}\Large)\cos(kz).\notag
	\end{align}   \label{EMquantization}\end{subequations} 
Here $\textbf{b}= \textbf{e}\times \textbf{k}/k$ is the magnetic field polarization, $\omega = k c$ is the optical frequency, and $U_{kn}(\textbf{r}) = e^{iz\nabla_\perp^2/k} u_n(x,y)$\footnote{\textcolor{black}{$U(\textbf{r})$ is a solution to the paraxial Helmholtz equation, and includes the Hermite-Gauss and Laguerre-Gauss modes. The initial condition $u(x,y)=U(x,y,0)$ corresponds to the tranverse modeshape function.}} \cite{wunsche2004quantization}, 
is a transverse-orthonormal basis for Gaussian beams satisfying
\begin{equation}\label{eq:orthonomality}
	\iint U_{kn}U^*_{k'n'}dxdy = \iint u_n u^*_{n'}dxdy  = \delta_{nn'},
\end{equation}
and $\hat{a}_n[\omega]$ is a bosonic operator with commutation relation $[\hat{a}_n[\omega],\hat{a}_{n'}^\dagger[\omega']] = \delta_{n,n'}\delta[\omega-\omega']$ normalized so that $\sum_n \int  \hat{a}_{n}^\dagger[\omega] \hat{a}_{n}[\omega] \t{d}\omega  $ and $\tfrac{1}{2\pi}\sum_n \iint  \hat{a}_{n}^\dagger[\omega] \hat{a}_{n}[\omega']e^{-i(\omega-\omega')t} \t{d}\omega\t{d}\omega'$ are the total photon number and flux, respectively \cite{blow1990continuum}.

A straightforward calculation using Eqs. \ref{EMquantization}-\ref{eq:orthonomality} and the rotating wave approximation (RWA) then yields
\begin{equation}
	\label{eq:HEMfree}
	\hat{H}_\t{EM}^\t{free} = \sum_{n}\hbar\int_0^\infty   \omega \hat{a}_{n}^\dagger(\omega)\hat{a}_{n}(\omega) \t{d}\omega
\end{equation}
and
\begin{equation}
	\hat{H}_{\mathrm{int}}\approx \hbar z_0\sum_{n,n'} \iint \frac{\sqrt{k k'}}{2\pi} \bigg(\hat{a}_{n}(\omega)\hat{a}^\dagger_{n'}[\omega']e^{-i(\omega-\omega')t} \beta_{n,n'} + \t{h.c.}\bigg)\t{d}\omega \t{d}\omega'
	\label{Hint1}
\end{equation}
where 
\color{black}\begin{equation}
	\beta_{n,n'}= \iint  u_n(x, y) u_{n'}^*(x, y) \phi(x, y) dxdy.
\end{equation}\color{black}

To simplify Eq. \ref{Hint1}, we assume that the laser has center frequency $\omega_0  = k_0 c$ and 
that interacting mode frequencies $\omega$ are confined to a narrow bandwidth $B\ll\omega_0$ such that $\sqrt{kk_0}\approx k_0$ and $\int_{0}^\infty\t{d}\omega\approx \int_{-\infty}^\infty\t{d}\omega$. 
Defining the time-dependent (\textcolor{black}{``Fourier-transformed'' }\footnote{\textcolor{black}{Eq. \ref{eq:spatialmodeoperator} is the inverse Fourier Transform (IFT) operator normalized ($1/\sqrt{2\pi}$) so that multimode Hamiltonians Eq. \ref{eq:HEMfree}, \ref{Hint1}, \ref{eq:OMHamiltonian} are integrated over angular frequencies, following a standard convention \cite{blow1990continuum,magrini2022squeezed,militaru2022}.  The Fourier transform defined in Eq. \ref{eq:FT}) uses a different IFT normalization ($1/(2\pi)$) so that PSDs are normalized over non-angular frequencies, $\langle X^2 \rangle = \int S_X(\omega)d\omega/(2\pi)$, following another standard convention \cite{aspelmeyerCavityOptomechanics2014,clerkIntroductionQuantumNoise2010}.}} \cite{blow1990continuum}) operator
\begin{equation}\label{eq:spatialmodeoperator}
	\hat{a}_n(t) = \frac{1}{\sqrt{2\pi}}\int_{-\infty}^\infty \hat{a}_n(\omega)e^{-i\omega t}\t{d}\omega
\end{equation} 
with commutation relation $[\hat{a}_n(t),\hat{a}_n(t')] = \delta (t-t')$, we then find (for notational simplicity, we hereafter use \textcolor{black}{$\hat{a}\equiv \hat{a}(t)$})
\begin{equation}\label{eq:Hint_2}
	H_{\mathrm{int}}
	\approx  \hbar k_0 z_0 \sum_{n,n'}\left(\beta_{n,n'}\hat{a}_n\hat{a}_{n'}^\dagger+\t{h.c.}\right).
\end{equation}

Equation \ref{eq:Hint_2} can be simplified further by choosing a spatial mode basis $\{u_n\}$ such that the laser mode is $u_\t{in} =u_0$. Expanding about this mode and keeping only cross terms $\beta_{n,n'}$ for which \textcolor{black}{$n$ or $n' = 0$} yields a principal mode expansion \cite{fabre2020modes}
\begin{equation}\label{eq:Hint_3}
	H_{\mathrm{int}}\approx
	2\hbar k_0 z_0\big(\beta_{0 ,0} \hat{a}^\dagger_0 \hat{a}_0 +\beta_\perp\big(\hat{a}_0^\dagger \hat{a}_\perp+\t{h.c.}\big)\big)
\end{equation}
where
\begin{equation}
	\hat{a}^\dagger_\perp = \beta_\perp^{-1}\sum_{n>0}\beta^\ast_{n ,0} \hat{a}^\dagger_{n}
\end{equation}
is an operator describing photon creation in the orthogonal spatial mode $u_\perp = \sum_{n>0}\beta_{n,0}u_n$ and $\beta_\perp^2 = \sum |\beta_{n,0}|^2$.

Going one step further, for each \textcolor{black}{Fourier-transformed} operator $\hat{a}_n(t)$ we can define temporally orthogonal components $\hat{a}_{n,\t{c}}(t)$ and $\hat{a}_{n,\t{m}}(t)$ whose frequency content is centered around the laser frequency $\omega_0$ and motional sideband frequencies $\omega_0\pm\omega_\t{m}$, respectively, with a bandwidth $B_\t{m}$ much smaller than the mechanical frequency $\omega_\t{m}$ and much larger than the mechanical linewidth \textcolor{black}{$\Gamma_\t{m}$}.  Inspecting Eq. \ref{Hint1} with $z_0(t) \approx z_0\cos(\omega_\t{m} t)$, it is evident the optomechanical interaction occurs between two spatiotemporally orthogonal modes (\textcolor{black}{with therefore commuting operators})
\begin{subequations}\begin{align}
		\hat{a}_\t{in}(t) &= \hat{a}_{0,\t{c}}(t) \textcolor{black}{,\; \t{and}} \label{eq:asc}\\
		\hat{a}_\t{sc}(t) &= \beta^{-1} \sum \beta_{0,n} \hat{a}_{0,\t{m}}(t)\\
		&\approx \beta^{-1}\left(\beta_{0,0} \hat{a}_{0,\t{m}}(t) + \beta_\perp a_\perp (t)\right)\\
		& \equiv \beta^{-1}\left(\beta_\parallel \hat{a}_\parallel (t)+ \beta_\perp \hat{a}_\perp(t)\right)
\end{align}\end{subequations}
with 
normalization $\beta^2 = \beta_\parallel^2 + \beta_\perp^2 = \sum_{n}|\beta_{0,n}|^2$.

Physically,  $\hat{a}_\t{sc}^\dagger$ describes creation of (temporally orthogonal) photons in spatial mode $u_\t{sc}$, and $\hat{a}^\dagger_\parallel$ and $\hat{a}^\dagger_\perp$ describe the fraction created in the incident spatial mode $u_\t{in} = u_0$ and its orthogonal complement $u_\perp$, respectively, defined by
\begin{equation}
	\begin{aligned}
		u_\t{sc} (x,y)=& \tunderbrace{\;\;\;\;\beta^{-1}}_{\textstyle \sqrt{\langle u_\t{in}\phi,u_\t{in}\phi\rangle^{-1}}} \tunderbrace{\sum_{ n }\beta_{n ,0} u_{n} (x,y)}_{\textstyle \phi u_\t{in}}, \\
		=&\beta^{-1}\big(\tunderbrace{\beta_{0,0}u_0(x,y)}_{\textstyle \beta_{\parallel}u_\parallel}+ \tunderbrace{\sum_{n >0}\beta_{n ,0} u_{n} (x,y)}_{\textstyle \beta_{\perp} u_\perp}\big)
	\end{aligned}
\end{equation} 

Rewriting Eq. \ref{eq:Hint_3} in terms of interacting modes with $z_0$ promoted to an operator $z_0\rightarrow \hat{z}$ \textcolor{black}{yields} Eq. 5 of the main text:
\begin{equation}\label{eq:HintQuantum_SI}
	\hat{H}_{\mathrm{int}}=2 \hbar k \beta \hat{z} \left(\hat{a}_\t{sc}\hat{a}_\t{in}^\dagger+\hat{a}_\t{sc}^\dagger\hat{a}_\t{in}\right)
\end{equation}

As a final simplification, we move to a frame rotating at the laser carrier frequency $\omega_0$ (by replacing $\hat{a}\rightarrow \hat{a} e^{i\omega_0 t}$) and assume that the input field is in a strong coherent state,
\begin{equation}\label{eq:coherentdrive}
	\hat{a}_\t{in}\rightarrow \sqrt{N}+\hat{a}_\t{in}
\end{equation}
where here $N\gg 1 $ is the incident photon flux and $\hat{a}_\t{in}$ has been redefined to represent vacuum fluctuations. Substituting Eq. \ref{eq:coherentdrive} into Eq. \ref{eq:HintQuantum_SI} and keeping only terms to first order in $\hat{a}_\t{in}$ yields the linearized interaction Hamiltonian used in Eq. 7
\begin{equation}
	\label{eq:HintAppendix}
	\hat{H}_{\mathrm{int}}= 2\hbar k  \sqrt{N} \beta\hat{X}_{\mathrm{sc}} \hat{z} = 2\hbar k  \sqrt{N}\left( \beta_\parallel\hat{X}_{\parallel}+\beta_\perp\hat{X}_{\perp}\right) \hat{z}
\end{equation}
where $\hat{X}= \hat{a}^\dagger+\hat{a}$
is the amplitude operator.

\vspace{-2mm}
\section{Optomechanical equation of motion, input-output relations, and quantum noise}
\vspace{-1mm}

In this section we derive the optomechanical equations of motion and quantum noise spectra discussed in the main text (Eqs. 8-12).  Our analysis proceeds from the linearized optomechanical Hamiltonian in the frame rotating at $\omega_0$ (Eqs. \ref{eq:coherentdrive}-\ref{eq:HintAppendix}), which, ignoring constants and bath terms, and operating in the \textcolor{black}{Schrödinger} picture (replacing $\hat{a}_n(\omega)e^{-i\omega t} \rightarrow \hat{a}_n(\omega,t)\equiv \hat{a}_n(\omega)$ in Eq. \ref{eq:spatialmodeoperator}), can be expressed as 
\begin{align*}\label{eq:OMHamiltonian}
	\numberthis
	\hat{H} &= \hat{H}_\t{EM}^\t{free} + \hat{H}_\t{m} + \hat{H}_\t{int} \\
	&\approx \int \hbar\omega \left( \hat{a}_\t{in}^\dagger(\omega) \hat{a}_\t{in}(\omega) + \hat{a}_\t{sc}^\dagger(\omega) \hat{a}_\t{sc}(\omega)\right) d\omega \\ &+ \hbar\omega_\t{m}\hat{b}^\dagger \hat{b}+ 2 \hbar k \sqrt{N}\beta \hat{z} \frac{1}{\sqrt{2\pi}}\int \left( \hat{a}_\t{sc}^\dagger(\omega) + \hat{a}_\t{sc}(\omega)\right)   d\omega
\end{align*}
where $\hat{b}$ is a bosonic operator for the mechanical oscillator satisfying $[\hat{b}, \hat{b}^\dagger]  = 1$, $\hat{z} = z_\t{ZP}(\hat{b}^\dagger + \hat{b})$ is the mechanical displacement operator, $z_\t{ZP} = \sqrt{\hbar/(2m_\t{eff}\omega_\t{m})}$ the oscillator's zero-point motion, and $m_\t{eff}$ is its effective mass.   

Equations of motion for $\hat{a}_\t{sc}(\omega)$ and $\hat{b}$ are obtained from the Heisenberg equation of motion $\dot{\hat{O}}(t) = (i/\hbar) [\hat{H},\hat{O}(t)]$:
\begin{subequations}\label{eq:HeisenbergEOMs}
	\begin{align}
		\dot{\hat{a}}_\t{sc}(\omega,t) & = -i \omega \hat{a}_\t{sc}(\omega,t) - 2 i k  \sqrt{N/(2\pi)}\beta\hat{z}(t) \\
		\label{eq:bEOM}
		\dot{\hat{b}}(t) &= - i \omega_\t{m} \hat{b}(t) - 2 i k \sqrt{N}\beta z_\t{ZP}\hat{X}_\t{sc}(t).
	\end{align}
\end{subequations}

\textcolor{black}{To find the field operators, we follow a standard procedure \cite{magrini2022squeezed, militaru2022}} and use the input-output relations \cite{gardinerInputOutputDamped1985} to treat the linear coupling between the input and reflected fields, yielding Eq.~7: 
\begin{equation}\label{eq:outputQuadratures}
	\hat{a}^\t{out}_\t{sc}(t) = \hat{a}^\t{in}_\t{sc}(t) + 2 i k \beta \sqrt{N} \hat{z}(t)  \\
\end{equation}
where $\hat{a}_\t{sc}$ (the output field operator) and $\hat{a}_\t{sc,0}$ (the input field operator) represent the scattered field before and after interacting with the membrane at time $t$, respectively.
Introducing the phase quadrature $\hat{Y}_\t{sc} = i(\hat{a}_\t{sc}-\hat{a}_\t{sc})$ and rotated quadrature $\hat{X}_\t{sc}^\theta = \hat{X}_\t{sc}\cos\theta+\hat{Y}_\t{sc}\sin\theta$ likewise yields 
\begin{equation}\label{eq:outputQuadratures2}
	\hat{X}_\t{sc}^{\theta,\t{out}}(t) = \hat{X}_\t{sc}^{\theta,\t{in}}(t) - 4 k \beta \sqrt{N} \hat{z}(t)\sin\theta. \\
\end{equation}
In the main text, we take $\hat{X}_\t{sc}^{\theta,\t{in}}$ to be vacuum noise.

Equation \ref{eq:outputQuadratures2} shows that displacement is encoded into the phase $(\theta = \pi/2)$ of the scattered field, while Eq. \ref{eq:HeisenbergEOMs}b 
suggests that backaction arises due to the scattered mode amplitude quantum noise.  
Substituting $z_\t{ZP}(\dot{\hat{b}}+\dot{\hat{b}}^\dagger) = \ddot{\hat{z}}$
yields the equation of motion for mechanical displacement,  
\begin{equation}
	\label{eq:mechEOM}
	m \ddot{\hat{z}}(t) + m \dot{\hat{z}}(t) \Gamma_\t{m} + m \omega_\t{m}^2 \hat{z}(t) = \hat{F}_\t{BA}(t)+ F_\t{th}(t)
\end{equation}
in which backaction manifests as a stochastic force 
\begin{equation}
	\hat{F}_\t{BA} = 2 \hbar k\sqrt{N}\beta\hat{X}^\t{in}_\t{sc}.
\end{equation}  
In Eq. \ref{eq:mechEOM} we have also introduced thermal noise via an ad-hoc dissipative force (characterized by mechanical damping rate $\Gamma_\t{m}$) and the associated thermal force $F_\t{th}$ due to the Fluctuation-Dissipation theorem \cite{saulsonThermalNoiseMechanical1990}.

In the frequency domain, Eqs. \ref{eq:outputQuadratures} and \ref{eq:mechEOM} constitute the optomechanical equations of motion in the main text (Eq. 8)
\begin{subequations}\begin{align}
		&	\hat{X}_\t{sc}^{\theta,\t{out}}(\omega) = \hat{X}_\t{sc}^{\theta,\t{in}}(\omega) - 4 k \beta \sqrt{N} z(\omega)\sin\theta \label{eq:Xscinoutfrequencydomain}\\
		&m\left( (\omega_\t{m}^2-\omega^2) + i\omega\Gamma_\t{m}\right) \hat{z}(\omega)  = \hat{F}_\t{BA}(\omega)+ F_\t{th}(\omega)\\&\hspace{39.5mm}\equiv \chi(\omega)^{-1}\hat{z}(\omega)
\end{align}\end{subequations}
where $\chi(\omega) = F(\omega)/z(\omega)$ is the mechanical susceptibility.

\textcolor{black}{We now compute the spectra of the optical quadratures and mechanical displacement, with the aim of deriving the generalized SQL given by Eq. 3 in the main text. Towards this end, we equate $S_z^\t{imp}$ in Eq. 3 with the apparent displacement PSD of an ideal homodyne measurement, defined as the displacement-equivalent PSD of the rotated field quadrature}
\begin{equation}\label{eq:Sztheta}
	S_z^\theta\equiv \left(16k^2\beta^2 N \sin^2\theta\right)^{-1}S^\t{out}_{X_\t{sc}^\theta}.
\end{equation}
where
\begin{equation}
	S_{X_\t{sc}^\theta} = \cos^2\theta S_{X_\t{sc}}+\sin^2\theta S_{Y_\t{sc}} +  \textcolor{black}{\sin(2\theta) \t{Re}\left[S_{X_\t{sc}Y_\t{sc}} \right]}
\end{equation}
\textcolor{black}{and here we have defined quantum noise PSD $S_X$ and CSD $S_{XY}$ as in Eqs. \ref{eq:SSPSD}-\ref{eq:SSCSD}, with $\{,\}$ and $\langle,\rangle$ corresponding to the anticommutator and expectation value, respectively \cite{clerkIntroductionQuantumNoise2010}.}

Expanding the right-hand side of Eq. \ref{eq:Sztheta} yields
\begin{subequations}\label{eq:SzApp}\begin{align}
		S_z^\theta&=\frac{S^\t{in}_{X_\t{sc}^\theta} }{16k^2\beta^2 N \sin^2\theta} + S_z + \textcolor{black}{{\frac{ \sin(2\theta) \t{Re}[S^\t{out}_{X_\t{sc}Y_\t{sc}}] }{\textcolor{black}{16}k^2\beta^2 N \sin^2\theta}}}\\
		&=\tunderbrace{\frac{S^\t{in}_{X_\t{sc}^\theta} }{16k^2\beta^2 N \sin^2\theta}}_{\textstyle S_z^\t{imp}} + \tunderbrace{|\chi|^2 S_F^\t{BA}}_{\textstyle S_z^\t{BA}}+\tunderbrace{|\chi|^2 S_F^\t{th}}_{\textstyle S_z^\t{th}} + \tunderbrace{\textcolor{black}{\frac{ \t{Re} [S^\t{out}_{X_\t{sc}Y_\t{sc}}] }{\textcolor{black}{8}k^2\beta^2 N\tan\theta}}}_{\textstyle S_z^\t{imp,BA}}
\end{align}\end{subequations}
where
$S_z^\t{imp}$, $S_z^\t{BA}$, $S_z^\t{th}$, and $S_z^\t{imp,BA}$ are the PSD of the apparent displacement (imprecision) due to shot noise, physical displacement due to backaction and thermal noise, and the correlation between imprecision and backaction noise, respectively.

Using Eq. \ref{eq:Xscinoutfrequencydomain} and the correlators $\langle \hat{a}_n (t) \hat{a}_m^\dagger(t') \rangle  = \delta_{\textcolor{black}{mn}}\delta (t-t')$ and $\langle \hat{a}_n^\dagger (t) \hat{a}_m(t') \rangle = 0$ yields
\begin{subequations}\label{eq:QuadNoise}
	\begin{align}
		S^\t{in}_{X^\theta_\t{sc}}(\omega) &= 2 \\
		S^\t{out}_{X_\t{sc}Y_\t{sc}}(\omega) &= - 16 \hbar k^2 \beta^2 N \chi (\omega)
	\end{align}
\end{subequations}
The non-zero CSD is due to the mechanical response to backaction, which couples amplitude and phase fluctuations. 

Collecting terms yields 
\begin{subequations}\label{eq:Appendix_SimpSbaSimpBA}
	\begin{align}
		S_z^\t{imp}& = (8 k^2 \beta^2 N \sin^2\theta)^{-1}\\
		S_z^\t{BA}&=8 \hbar^2 k^2 \beta^2 N |\chi|^2 = |\chi|^2 S_F^\t{BA}\\
		S_z^{\t{imp,BA}}& = \textcolor{black}{-\textcolor{black}{2} \hbar \cot\theta \t{Re}[\chi]}
	\end{align}
\end{subequations}
recovering expressions in the main text.  Finally, combining Eqs. \ref{eq:Appendix_SimpSbaSimpBA}a,b yields Eq. 3 of the main text: $S_z^\t{imp}S_F^\t{BA} = \hbar^2$.

\section{Ponderomotive Entanglement}

We conclude by deriving Eq. 12 in the main text, describing quantum correlations between the two orthogonal spatial modes comprising the scattered field. To this end, we substitute $\hat{a}_\t{sc} = \beta^{-1}(\beta_\parallel \hat{a}_\parallel + \beta_\perp \hat{a}_\perp)$ into Eq. \ref{eq:OMHamiltonian} and apply an analogous procedure to obtain input-output relations
\begin{subequations}\begin{align}
		\hat{X}_\perp^{\theta,\t{out}}(t)& = \hat{X}_{\perp}^{\theta,\t{in}}(t) - 4 k_0 \beta_\perp \sqrt{N} z(t)\sin\theta  \\
		\hat{X}_{\parallel}^{\theta,\t{out}}(t) & = \hat{X}_{\parallel}^{\theta,\t{in}}(t) - 4 k_0 \beta_\parallel \sqrt{N} z(t)\sin\theta 
\end{align}\end{subequations}
and noise spectra
\begin{subequations}\begin{align}
		S^\t{in}_{X_{\perp(\parallel)}^\theta}(\omega) & = 2\\
		S^\t{out}_{X_{\perp(\parallel)}Y_{\perp(\parallel)}}(\omega) & = - 16 \hbar k^2 \beta_{\perp(\parallel)}^2 N \chi (\omega)\\
  S^\t{out}_{X_\parallel Y_\perp}(\omega) &= -16 \hbar k^2 \beta_\perp \beta_\parallel N \chi (\omega)\label{eq:twomodeentanglement_SI}.
\end{align}\end{subequations}

\begin{figure}[b!]
	\centering  
	\includegraphics[width=0.8\columnwidth]{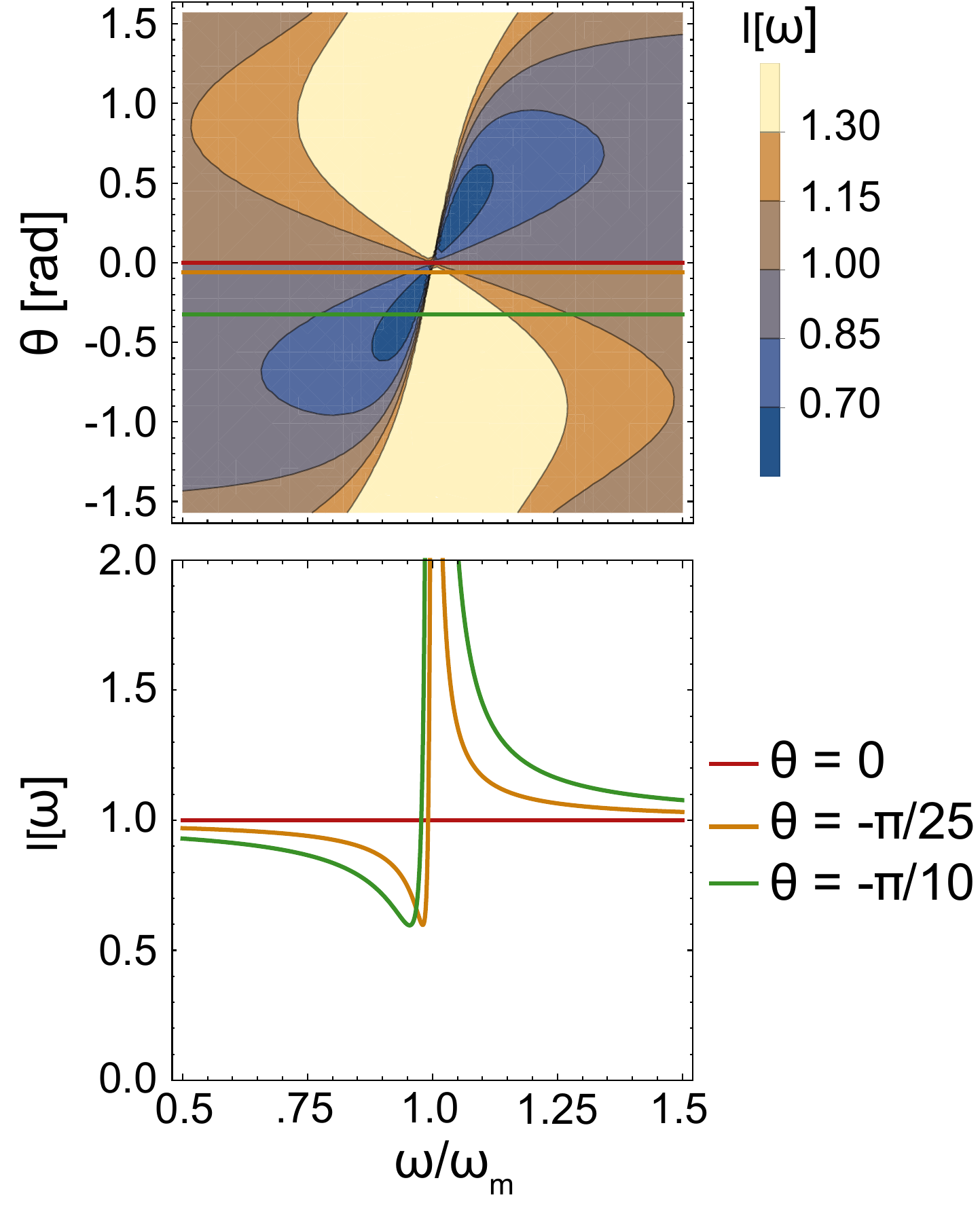}
	\caption{(a) DGCZ criterion $I$ (Eq. \ref{eq:IImportant}) versus Fourier frequency $\omega$ and homodyne phase angle $\theta$, with $\omega_\t{m} = 2\pi \times 40 \: \t{kHz},$ $m = 10^{-12}$~kg,  $\omega_\t{m}/\Gamma_\t{m} = 4 \times 10^7$, $k^2 \beta^2 N = 2 \pi^2 \times 10^{24} \: \t{m}^{-2} \: \t{s}^{-1}$, 
    and zero bath temperature $(S_z^\t{th} = 2 \hbar \omega_\t{m} \Gamma_\t{m} m |\chi|^2$). (b) Cuts along three phase angles, corresponding to horizontal lines in (a). }
	\label{fig:FigS1} 
	\vspace{-2mm}
\end{figure}

The correlations given by Eq. \ref{eq:twomodeentanglement_SI} are a signature of ``ponderomotive entanglement'' \cite{chenEntanglementPropagatingOptical2020}---i.e., correlations between two otherwise independent (in this case spatially orthogonal) optical modes due to their mutual coupling to a mechanical oscillator. To verify entanglement, one can construct EPR-like quadratures of the joint system $\hat{X}_+ = \hat{X}_\parallel^\theta + \hat{X}_\perp^\theta$ and $\hat{Y}_- = \hat{Y}_\parallel^\theta - \hat{Y}_\perp^\theta$, where $\hat{Y}^\theta = \sin(\theta) \hat{X} + \cos(\theta) \hat{Y}$. Since the linearized interaction preserves the Gaussianity of its input states, a necessary and sufficient condition for entanglement is the Duan–Giedke–Cirac–Zoller (DGCZ) criterion~\cite{duan2000inseparability, fabre2020modes, chenEntanglementPropagatingOptical2020},  
\begin{equation}
	\label{eq:DGCZCritFreq}
	I (\omega) \equiv \frac{S^\t{out}_{X_+}(\omega) + S^\t{out}_{Y_-} (\omega)}{8} < 1,
\end{equation}
which here we have normalized to the vacuum state single-sided PSD of 2. 
Considering the simplified scenario $\bar{\beta} = \beta_\parallel = \beta_\perp \neq 0$ and dual homodyne measurements performed on both modes at equal quadrature measurement angles $\theta$, we find
\begin{subequations}\begin{align}
    	\label{eq:IImportant}
	I (\omega) &= 1 + \textcolor{black}{8} k^2 \bar{\beta}^2 N |\chi|^2 (S_F^\t{BA, \parallel} +  S_F^\t{BA, \perp} + S_F^\t{th}) \sin^2(\theta) \notag\\
	\numberthis
	&\;\;\;\;\;\;\;\;-  \textcolor{black}{8} \hbar k^2 \bar{\beta}^2 N \t{Re}[\chi] \sin(2\theta)\\
 & = 1 + \textcolor{black}{8} k^2 \bar{\beta}^2 N \sin^2\theta\left(S_z^\t{BA} + S_z^\t{th} + S_z^\t{imp,BA}/2\right)
\end{align}\end{subequations}
Where $S_F^\t{BA, \parallel(\perp)} = (\beta_{\parallel(\perp)}/\beta)^2 S_F^\t{BA}$ is the fraction of the backaction force due to temporal (spatial) photon shot noise. For small quadrature angles $\theta$ in the backaction-dominated regime ($S_z^\t{BA} > S_z^\t{th}$), the correlation term $S_z^\t{imp,BA}$ produces $I (\omega) < 1$ near the mechanical resonance.  
 Representative plots of the DGCZ criterion for different $\theta$ are shown in Fig. \ref{fig:FigS1}.


}

\end{document}